\documentclass[conference]{IEEEtran}
\IEEEoverridecommandlockouts
\usepackage{cite}
\usepackage{amsmath,amssymb,amsfonts}
\usepackage{algorithmic}
\usepackage{graphicx}
\usepackage{subcaption}
\usepackage{textcomp}
\usepackage{enumitem}
\usepackage{xcolor}
\def\BibTeX{{\rm B\kern-.05em{\sc i\kern-.025em b}\kern-.08em
    T\kern-.1667em\lower.7ex\hbox{E}\kern-.125emX}}

\usepackage{fancyvrb}

\makeatletter
\def\PYG@reset{\let\PYG@it=\relax \let\PYG@bf=\relax%
    \let\PYG@ul=\relax \let\PYG@tc=\relax%
    \let\PYG@bc=\relax \let\PYG@ff=\relax}
\def\PYG@tok#1{\csname PYG@tok@#1\endcsname}
\def\PYG@toks#1+{\ifx\relax#1\empty\else%
    \PYG@tok{#1}\expandafter\PYG@toks\fi}
\def\PYG@do#1{\PYG@bc{\PYG@tc{\PYG@ul{%
    \PYG@it{\PYG@bf{\PYG@ff{#1}}}}}}}
\def\PYG#1#2{\PYG@reset\PYG@toks#1+\relax+\PYG@do{#2}}

\@namedef{PYG@tok@w}{\def\PYG@tc##1{\textcolor[rgb]{0.73,0.73,0.73}{##1}}}
\@namedef{PYG@tok@c}{\def\PYG@tc##1{\textcolor[rgb]{0.53,0.53,0.53}{##1}}}
\@namedef{PYG@tok@cp}{\let\PYG@bf=\textbf\def\PYG@tc##1{\textcolor[rgb]{0.80,0.00,0.00}{##1}}}
\@namedef{PYG@tok@cs}{\let\PYG@bf=\textbf\def\PYG@tc##1{\textcolor[rgb]{0.80,0.00,0.00}{##1}}\def\PYG@bc##1{{\setlength{\fboxsep}{0pt}\colorbox[rgb]{1.00,0.94,0.94}{\strut ##1}}}}
\@namedef{PYG@tok@s}{\def\PYG@tc##1{\textcolor[rgb]{0.87,0.13,0.00}{##1}}\def\PYG@bc##1{{\setlength{\fboxsep}{0pt}\colorbox[rgb]{1.00,0.94,0.94}{\strut ##1}}}}
\@namedef{PYG@tok@sr}{\def\PYG@tc##1{\textcolor[rgb]{0.00,0.53,0.00}{##1}}\def\PYG@bc##1{{\setlength{\fboxsep}{0pt}\colorbox[rgb]{1.00,0.94,1.00}{\strut ##1}}}}
\@namedef{PYG@tok@sx}{\def\PYG@tc##1{\textcolor[rgb]{0.13,0.73,0.13}{##1}}\def\PYG@bc##1{{\setlength{\fboxsep}{0pt}\colorbox[rgb]{0.94,1.00,0.94}{\strut ##1}}}}
\@namedef{PYG@tok@ss}{\def\PYG@tc##1{\textcolor[rgb]{0.67,0.40,0.00}{##1}}\def\PYG@bc##1{{\setlength{\fboxsep}{0pt}\colorbox[rgb]{1.00,0.94,0.94}{\strut ##1}}}}
\@namedef{PYG@tok@si}{\def\PYG@tc##1{\textcolor[rgb]{0.20,0.20,0.73}{##1}}\def\PYG@bc##1{{\setlength{\fboxsep}{0pt}\colorbox[rgb]{1.00,0.94,0.94}{\strut ##1}}}}
\@namedef{PYG@tok@se}{\def\PYG@tc##1{\textcolor[rgb]{0.00,0.27,0.87}{##1}}\def\PYG@bc##1{{\setlength{\fboxsep}{0pt}\colorbox[rgb]{1.00,0.94,0.94}{\strut ##1}}}}
\@namedef{PYG@tok@ow}{\def\PYG@tc##1{\textcolor[rgb]{0.00,0.53,0.00}{##1}}}
\@namedef{PYG@tok@k}{\let\PYG@bf=\textbf\def\PYG@tc##1{\textcolor[rgb]{0.00,0.53,0.00}{##1}}}
\@namedef{PYG@tok@kp}{\def\PYG@tc##1{\textcolor[rgb]{0.00,0.53,0.00}{##1}}}
\@namedef{PYG@tok@kt}{\let\PYG@bf=\textbf\def\PYG@tc##1{\textcolor[rgb]{0.53,0.53,0.53}{##1}}}
\@namedef{PYG@tok@nc}{\let\PYG@bf=\textbf\def\PYG@tc##1{\textcolor[rgb]{0.73,0.00,0.40}{##1}}}
\@namedef{PYG@tok@ne}{\let\PYG@bf=\textbf\def\PYG@tc##1{\textcolor[rgb]{0.73,0.00,0.40}{##1}}}
\@namedef{PYG@tok@nf}{\let\PYG@bf=\textbf\def\PYG@tc##1{\textcolor[rgb]{0.00,0.40,0.73}{##1}}}
\@namedef{PYG@tok@py}{\let\PYG@bf=\textbf\def\PYG@tc##1{\textcolor[rgb]{0.20,0.40,0.60}{##1}}}
\@namedef{PYG@tok@nn}{\let\PYG@bf=\textbf\def\PYG@tc##1{\textcolor[rgb]{0.73,0.00,0.40}{##1}}}
\@namedef{PYG@tok@nb}{\def\PYG@tc##1{\textcolor[rgb]{0.00,0.20,0.53}{##1}}}
\@namedef{PYG@tok@nv}{\def\PYG@tc##1{\textcolor[rgb]{0.20,0.40,0.60}{##1}}}
\@namedef{PYG@tok@vc}{\def\PYG@tc##1{\textcolor[rgb]{0.20,0.40,0.60}{##1}}}
\@namedef{PYG@tok@vi}{\def\PYG@tc##1{\textcolor[rgb]{0.20,0.20,0.73}{##1}}}
\@namedef{PYG@tok@vg}{\def\PYG@tc##1{\textcolor[rgb]{0.87,0.47,0.00}{##1}}}
\@namedef{PYG@tok@no}{\let\PYG@bf=\textbf\def\PYG@tc##1{\textcolor[rgb]{0.00,0.20,0.40}{##1}}}
\@namedef{PYG@tok@nt}{\let\PYG@bf=\textbf\def\PYG@tc##1{\textcolor[rgb]{0.73,0.00,0.40}{##1}}}
\@namedef{PYG@tok@na}{\def\PYG@tc##1{\textcolor[rgb]{0.20,0.40,0.60}{##1}}}
\@namedef{PYG@tok@nd}{\def\PYG@tc##1{\textcolor[rgb]{0.33,0.33,0.33}{##1}}}
\@namedef{PYG@tok@nl}{\let\PYG@it=\textit\def\PYG@tc##1{\textcolor[rgb]{0.20,0.40,0.60}{##1}}}
\@namedef{PYG@tok@m}{\let\PYG@bf=\textbf\def\PYG@tc##1{\textcolor[rgb]{0.00,0.00,0.87}{##1}}}
\@namedef{PYG@tok@gh}{\def\PYG@tc##1{\textcolor[rgb]{0.20,0.20,0.20}{##1}}}
\@namedef{PYG@tok@gu}{\def\PYG@tc##1{\textcolor[rgb]{0.40,0.40,0.40}{##1}}}
\@namedef{PYG@tok@gd}{\def\PYG@tc##1{\textcolor[rgb]{0.00,0.00,0.00}{##1}}\def\PYG@bc##1{{\setlength{\fboxsep}{0pt}\colorbox[rgb]{1.00,0.87,0.87}{\strut ##1}}}}
\@namedef{PYG@tok@gi}{\def\PYG@tc##1{\textcolor[rgb]{0.00,0.00,0.00}{##1}}\def\PYG@bc##1{{\setlength{\fboxsep}{0pt}\colorbox[rgb]{0.87,1.00,0.87}{\strut ##1}}}}
\@namedef{PYG@tok@gr}{\def\PYG@tc##1{\textcolor[rgb]{0.67,0.00,0.00}{##1}}}
\@namedef{PYG@tok@ge}{\let\PYG@it=\textit}
\@namedef{PYG@tok@gs}{\let\PYG@bf=\textbf}
\@namedef{PYG@tok@gp}{\def\PYG@tc##1{\textcolor[rgb]{0.33,0.33,0.33}{##1}}}
\@namedef{PYG@tok@go}{\def\PYG@tc##1{\textcolor[rgb]{0.53,0.53,0.53}{##1}}}
\@namedef{PYG@tok@gt}{\def\PYG@tc##1{\textcolor[rgb]{0.67,0.00,0.00}{##1}}}
\@namedef{PYG@tok@err}{\def\PYG@tc##1{\textcolor[rgb]{0.65,0.09,0.09}{##1}}\def\PYG@bc##1{{\setlength{\fboxsep}{0pt}\colorbox[rgb]{0.89,0.82,0.82}{\strut ##1}}}}
\@namedef{PYG@tok@kc}{\let\PYG@bf=\textbf\def\PYG@tc##1{\textcolor[rgb]{0.00,0.53,0.00}{##1}}}
\@namedef{PYG@tok@kd}{\let\PYG@bf=\textbf\def\PYG@tc##1{\textcolor[rgb]{0.00,0.53,0.00}{##1}}}
\@namedef{PYG@tok@kn}{\let\PYG@bf=\textbf\def\PYG@tc##1{\textcolor[rgb]{0.00,0.53,0.00}{##1}}}
\@namedef{PYG@tok@kr}{\let\PYG@bf=\textbf\def\PYG@tc##1{\textcolor[rgb]{0.00,0.53,0.00}{##1}}}
\@namedef{PYG@tok@bp}{\def\PYG@tc##1{\textcolor[rgb]{0.00,0.20,0.53}{##1}}}
\@namedef{PYG@tok@fm}{\let\PYG@bf=\textbf\def\PYG@tc##1{\textcolor[rgb]{0.00,0.40,0.73}{##1}}}
\@namedef{PYG@tok@vm}{\def\PYG@tc##1{\textcolor[rgb]{0.20,0.40,0.60}{##1}}}
\@namedef{PYG@tok@sa}{\def\PYG@tc##1{\textcolor[rgb]{0.87,0.13,0.00}{##1}}\def\PYG@bc##1{{\setlength{\fboxsep}{0pt}\colorbox[rgb]{1.00,0.94,0.94}{\strut ##1}}}}
\@namedef{PYG@tok@sb}{\def\PYG@tc##1{\textcolor[rgb]{0.87,0.13,0.00}{##1}}\def\PYG@bc##1{{\setlength{\fboxsep}{0pt}\colorbox[rgb]{1.00,0.94,0.94}{\strut ##1}}}}
\@namedef{PYG@tok@sc}{\def\PYG@tc##1{\textcolor[rgb]{0.87,0.13,0.00}{##1}}\def\PYG@bc##1{{\setlength{\fboxsep}{0pt}\colorbox[rgb]{1.00,0.94,0.94}{\strut ##1}}}}
\@namedef{PYG@tok@dl}{\def\PYG@tc##1{\textcolor[rgb]{0.87,0.13,0.00}{##1}}\def\PYG@bc##1{{\setlength{\fboxsep}{0pt}\colorbox[rgb]{1.00,0.94,0.94}{\strut ##1}}}}
\@namedef{PYG@tok@sd}{\def\PYG@tc##1{\textcolor[rgb]{0.87,0.13,0.00}{##1}}\def\PYG@bc##1{{\setlength{\fboxsep}{0pt}\colorbox[rgb]{1.00,0.94,0.94}{\strut ##1}}}}
\@namedef{PYG@tok@s2}{\def\PYG@tc##1{\textcolor[rgb]{0.87,0.13,0.00}{##1}}\def\PYG@bc##1{{\setlength{\fboxsep}{0pt}\colorbox[rgb]{1.00,0.94,0.94}{\strut ##1}}}}
\@namedef{PYG@tok@sh}{\def\PYG@tc##1{\textcolor[rgb]{0.87,0.13,0.00}{##1}}\def\PYG@bc##1{{\setlength{\fboxsep}{0pt}\colorbox[rgb]{1.00,0.94,0.94}{\strut ##1}}}}
\@namedef{PYG@tok@s1}{\def\PYG@tc##1{\textcolor[rgb]{0.87,0.13,0.00}{##1}}\def\PYG@bc##1{{\setlength{\fboxsep}{0pt}\colorbox[rgb]{1.00,0.94,0.94}{\strut ##1}}}}
\@namedef{PYG@tok@mb}{\let\PYG@bf=\textbf\def\PYG@tc##1{\textcolor[rgb]{0.00,0.00,0.87}{##1}}}
\@namedef{PYG@tok@mf}{\let\PYG@bf=\textbf\def\PYG@tc##1{\textcolor[rgb]{0.00,0.00,0.87}{##1}}}
\@namedef{PYG@tok@mh}{\let\PYG@bf=\textbf\def\PYG@tc##1{\textcolor[rgb]{0.00,0.00,0.87}{##1}}}
\@namedef{PYG@tok@mi}{\let\PYG@bf=\textbf\def\PYG@tc##1{\textcolor[rgb]{0.00,0.00,0.87}{##1}}}
\@namedef{PYG@tok@il}{\let\PYG@bf=\textbf\def\PYG@tc##1{\textcolor[rgb]{0.00,0.00,0.87}{##1}}}
\@namedef{PYG@tok@mo}{\let\PYG@bf=\textbf\def\PYG@tc##1{\textcolor[rgb]{0.00,0.00,0.87}{##1}}}
\@namedef{PYG@tok@ch}{\def\PYG@tc##1{\textcolor[rgb]{0.53,0.53,0.53}{##1}}}
\@namedef{PYG@tok@cm}{\def\PYG@tc##1{\textcolor[rgb]{0.53,0.53,0.53}{##1}}}
\@namedef{PYG@tok@cpf}{\def\PYG@tc##1{\textcolor[rgb]{0.53,0.53,0.53}{##1}}}
\@namedef{PYG@tok@c1}{\def\PYG@tc##1{\textcolor[rgb]{0.53,0.53,0.53}{##1}}}

\def\PYGZus{\char`\_}

\def\PYGZhy{\char`\-}

\makeatother

\begin{document}

\title{An Adaptive Distributed Stencil Abstraction for GPUs
}

\author{\IEEEauthorblockN{Aditya Bhosale}
\IEEEauthorblockA{\textit{University of Illinois at Urbana-Champaign}\\
Urbana, IL, USA \\
adityapb@illinois.edu}
\and
\IEEEauthorblockN{Laxmikant Kale}
\IEEEauthorblockA{\textit{University of Illinois at Urbana-Champaign}\\
Urbana, IL, USA \\
kale@illinois.edu}
}

\IEEEpubid{\makebox[\columnwidth]{\parbox{\columnwidth}{
    \vspace*{1.27cm} % <-- ADJUST THIS VALUE TO ADD MORE EMPTY SPACE
    \hfill
}} \hspace{\columnsep}\makebox[\columnwidth]{ }}

\maketitle

\begin{abstract}
The scientific computing ecosystem in Python is largely confined to single-node parallelism, creating a gap between high-level prototyping in NumPy and high-performance execution on modern supercomputers. The increasing prevalence of hardware accelerators and the need for energy efficiency have made resource adaptivity a critical requirement, yet traditional HPC abstractions remain rigid. To address these challenges, we present an adaptive, distributed abstraction for stencil computations on multi-node GPUs. This abstraction is built using CharmTyles, a framework based on the adaptive Charm++ runtime, and features a familiar NumPy-like syntax to minimize the porting effort from prototype to production code. 
We showcase the resource elasticity of our abstraction by dynamically rescaling a running application across a different number of nodes and present a performance analysis of the associated overheads.
Furthermore, we demonstrate that our abstraction achieves significant performance improvements over both a specialized, high-performance stencil DSL and a generalized NumPy replacement. 

\end{abstract}


\section{Introduction}

Python has emerged as a language of choice for scientific computing due to its large ecosystem of composable libraries built upon NumPy's common array representation~\cite{numpy}. In particular, NumPy's slicing notation is widely used for implementing numerical methods, such as finite difference and finite volume, over structured grids. This pattern of stencil computations has a wide range of applications, including computational fluid dynamics, seismic imaging, electromagnetism, and image processing. While NumPy exhibits good single-threaded performance by internally calling highly efficient BLAS functions and supports multi-threaded execution for certain operations, its parallelism is limited to a single node.

Over the last few decades, parallel programming has seen a steady shift from shared-memory architectures to distributed-memory systems and now to hardware accelerators like GPUs and TPUs. Each new hardware architecture introduces its own programming model, optimization challenges, and performance constraints. This changing landscape of parallel programming has made it difficult for domain scientists to effectively utilize available resources. Domain-specific languages (DSLs) are a powerful tool to separate domain expertise from the technical knowledge required to extract optimal performance from different hardware architectures. To that end, several Python-based DSLs have been developed for high-performance stencil computations~\cite{pystencils, pystella, devito-compiler}.

A common workflow in scientific computing is to first build a prototype using NumPy to evaluate an algorithm on a smaller scale and then develop a high-performance implementation. Unfortunately, since these high-level DSLs often differ significantly in syntax from NumPy, the effort required to port the implementation is substantial. Several NumPy drop-in replacements have been developed to tackle this issue~\cite{cupynumeric, triton}; however, these replacements often compromise on domain-specific performance optimization in the pursuit of generality.

Due to the rise in popularity of AI, the demand for accelerators has increased exponentially over the past few years. This sudden rise in demand has also strained the energy requirements for datacenters and supercomputers, making it more important than ever to use these resources efficiently. Moreover, as issues like hardware failures and power constraints become more pertinent in modern supercomputing systems, it is crucial to introduce adaptivity and resource awareness into job schedulers and parallel programming models. The widespread adoption of cloud platforms for HPC workloads has underscored the importance of resilience, fault-tolerance, and resource elasticity in parallel programming models and abstractions to effectively use these volatile cloud resources.

Traditionally, parallel programming models like MPI have been rigid. As a result, job schedulers like SLURM have not supported dynamic allocation and adaptivity. In recent years, significant work has been done to develop alternative, adaptive parallel programming models such as Charm++~\cite{malleable2014}, and to extend MPI to incorporate elasticity~\cite{mpi2, mpisessions}. Similarly, there has been work on introducing adaptivity to traditional job schedulers like SLURM~\cite{slurm-extend} and exploring resource elasticity in the HPC-Cloud converged computing paradigm~\cite{flux, Kub2023, Bhosale2025Cloud, BhosaleCanopie25}. While several popular libraries for distributed machine learning have adapted to this changing trend by introducing elasticity into their abstractions~\cite{elasticflow}, traditional HPC libraries and DSLs have largely continued to be rigid.

CharmTyles is a framework based on the Charm++ programming model for writing DSLs with a Python frontend and a parallel Charm++ backend~\cite{charmstencil}. In this paper, we extend the resource elasticity in Charm++ and CharmTyles to support GPUs. We present an adaptive, distributed abstraction for stencil computations on multi-node GPUs built using CharmTyles with a NumPy-like syntax. We compare the performance of our abstraction to Devito~\cite{devito-compiler}, a specialized stencil DSL, and cuPyNumeric~\cite{cupynumeric}, a generalized NumPy replacement, showing a significant performance improvement over both. We also demonstrate the adaptivity of our abstraction by dynamically rescaling an application to a different number of nodes, and we measure the performance overheads associated with this rescaling.

\section{Background}

\subsection{Charm++}

Charm++ is an asynchronous message-driven parallel programming model~\cite{sc14charm} where computation is expressed in terms of objects known as \textit{chares}. The Charm++ runtime maps these chares to Processing Elements (PEs) and manages communication between them through remote entry method invocations. A key feature of Charm++ is that its chares are migratable, allowing the runtime system to move them between PEs to perform dynamic load balancing and support resource elasticity.

\subsubsection{GPU support}

GPU support in Charm++ is implemented using the Hybrid API (HAPI)~\cite{CharmHAPI}. HAPI provides three main functionalities:

\paragraph{Mapping PEs to GPUs} 
Modern supercomputers have many CPU cores but few GPUs per node, so several PEs must share a single GPU. In Charm++, HAPI manages this PE-to-GPU mapping using built-in or custom, user-defined functions.

\paragraph{Asynchronous detection of kernel completion} 
Charm++ programs often use a continuation-passing style. Due to its message-driven model, an entry method performs local computation and then invokes a remote method to continue. This creates a dependency between the local computation and the subsequent remote invocation.

If the local computation is a CUDA kernel, enforcing this dependency would require a blocking call, which is suboptimal as it can delay other messages in the PE's scheduler. HAPI provides a callback to resolve this: the kernel call is made asynchronous, and a callback containing the subsequent message is registered. The Charm++ runtime delivers this callback only after the kernel is complete.

\paragraph{Direct GPU messaging}

HAPI supports sending GPU buffers as entry method arguments using the Charm++ zero-copy API. Communication of GPU buffers is performed using GPUDirect RDMA when Charm++ is built with the UCX communication layer~\cite{CharmGPUUCX}. For other communication layers, intra-node GPU transfers are done using CUDA IPC and CUDA memcpy, while inter-node GPU communication uses host-staging.

\subsubsection{Shrink/Expand support}

The migratability of chares enables an application to dynamically scale its number of PEs up or down at runtime. When scaling down (shrinking), chares located on the PEs designated for removal are migrated to the remaining PEs. Conversely, when scaling up (expanding), chares from existing PEs are moved to the newly added PEs.

To manage the internal data structure modifications required by rescaling, the application is checkpointed locally and restarted with the new number of PEs.
In a shrink operation, chares are first migrated away from the PEs being removed. The application is then checkpointed to shared memory, restarted with fewer PEs, and restored.
During an expand operation, the application is first checkpointed and restarted with the new PEs. After the checkpoint data is restored, a load balancing procedure is executed to migrate chares onto the new PEs.

A signal for an application to rescale can be sent from an external program using the Converse Client Server (CCS) interface~\cite{charm-ccs}. The rescaling operation is then triggered during the application's subsequent load balancing step.

\subsection{CharmTyles}

\begin{figure}[!ht]
    \centering
    \includegraphics[width=\linewidth]{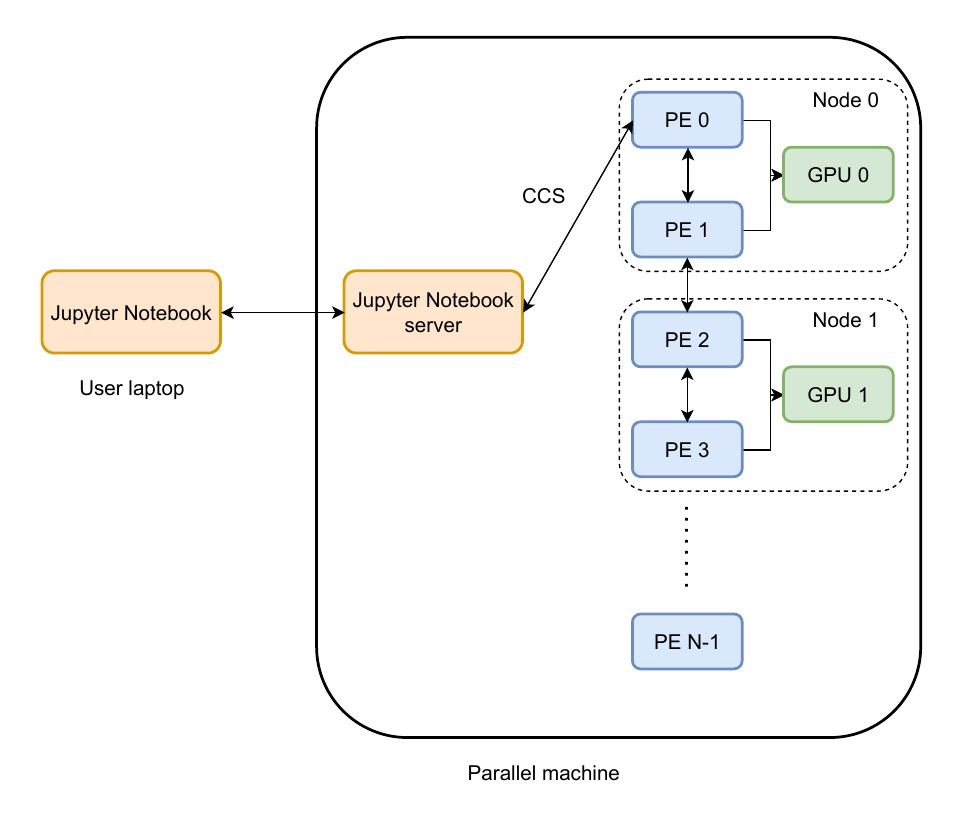}
    \caption{CharmTyles architecture with GPUs}
    \label{fig:charmtyles}
\end{figure}

CharmTyles is a framework for writing DSLs using a client-server architecture with a Python frontend and a Charm++ backend~\cite{charmstencil}. As shown in Figure~\ref{fig:charmtyles}, the framework consists of a Charm++ server program running on the backend. When running on GPU systems, this program uses PEs that are processes or threads running on the CPU, with each PE mapped to a GPU device. Data on the backend is distributed across these PEs as \textit{tiles}.

The frontend is a Python interpreter where the user expresses computation in a high-level DSL. The frontend encodes the computation into a message that is sent via CCS to the backend, where the Charm++ program executes it.

CharmStencil is a distributed stencil abstraction for CPUs that was previously built using the CharmTyles framework~\cite{charmstencil}. In this paper, we extend CharmStencil to support execution on GPUs and modify the frontend to more closely resemble NumPy's syntax. Henceforth in this paper, ``CharmStencil'' will refer to the new GPU implementation of this abstraction developed as part of this work.

\section{Implementation}

\subsection{Overview}

We implement our stencil abstraction using the CharmTyles framework. On the frontend, the user writes operations with a NumPy-like syntax. These operations are aggregated and sent to the backend, creating a pipelined execution model that helps hide the overhead introduced by the Python interpreter.

Example~\ref{code:jacobi} shows a 2D Laplace equation solver written using CharmStencil, which has a syntax very similar to a NumPy implementation. A notable difference is that the CharmStencil implementation requires two arrays, \texttt{u1} and \texttt{u2}, whereas the same application in NumPy could be written with a single array. This design choice avoids the creation of temporary arrays for intermediate results, instead requiring the user to write expressions without data dependencies. The frontend will detect any such dependencies and raise an error.

\begin{figure}[!ht]
\centering
\begin{Verbatim}[fontsize=\footnotesize, commandchars=\\\{\}]
\PYG{n}{u1} \PYG{o}{=} \PYG{n}{create\PYGZus{}array}\PYG{p}{((}\PYG{l+m+mi}{16384}\PYG{p}{,} \PYG{l+m+mi}{16384}\PYG{p}{))}
\PYG{n}{u2} \PYG{o}{=} \PYG{n}{create\PYGZus{}array}\PYG{p}{((}\PYG{l+m+mi}{16384}\PYG{p}{,} \PYG{l+m+mi}{16384}\PYG{p}{))}

\PYG{n}{u1}\PYG{p}{[}\PYG{l+m+mi}{0}\PYG{p}{,} \PYG{p}{:]} \PYG{o}{=} \PYG{n}{u1}\PYG{p}{[}\PYG{o}{\PYGZhy{}}\PYG{l+m+mi}{1}\PYG{p}{,} \PYG{p}{:]} \PYG{o}{=} \PYG{n}{u1}\PYG{p}{[:,} \PYG{l+m+mi}{0}\PYG{p}{]} \PYG{o}{=} \PYG{n}{u1}\PYG{p}{[:,} \PYG{o}{\PYGZhy{}}\PYG{l+m+mi}{1}\PYG{p}{]} \PYG{o}{=} \PYG{l+m+mi}{1}
\PYG{n}{u2}\PYG{p}{[}\PYG{l+m+mi}{0}\PYG{p}{,} \PYG{p}{:]} \PYG{o}{=} \PYG{n}{u2}\PYG{p}{[}\PYG{o}{\PYGZhy{}}\PYG{l+m+mi}{1}\PYG{p}{,} \PYG{p}{:]} \PYG{o}{=} \PYG{n}{u2}\PYG{p}{[:,} \PYG{l+m+mi}{0}\PYG{p}{]} \PYG{o}{=} \PYG{n}{u2}\PYG{p}{[:,} \PYG{o}{\PYGZhy{}}\PYG{l+m+mi}{1}\PYG{p}{]} \PYG{o}{=} \PYG{l+m+mi}{1}

\PYG{k}{for} \PYG{n}{i} \PYG{o+ow}{in} \PYG{n+nb}{range}\PYG{p}{(}\PYG{l+m+mi}{10}\PYG{p}{):}
    \PYG{n}{u2}\PYG{p}{[}\PYG{l+m+mi}{1}\PYG{p}{:}\PYG{o}{\PYGZhy{}}\PYG{l+m+mi}{1}\PYG{p}{,} \PYG{l+m+mi}{1}\PYG{p}{:}\PYG{o}{\PYGZhy{}}\PYG{l+m+mi}{1}\PYG{p}{]} \PYG{o}{=} \PYG{l+m+mf}{0.25} \PYG{o}{*} \PYG{p}{(}\PYG{n}{u1}\PYG{p}{[:}\PYG{o}{\PYGZhy{}}\PYG{l+m+mi}{2}\PYG{p}{,} \PYG{l+m+mi}{1}\PYG{p}{:}\PYG{o}{\PYGZhy{}}\PYG{l+m+mi}{1}\PYG{p}{]} \PYG{o}{+}
                             \PYG{n}{u1}\PYG{p}{[}\PYG{l+m+mi}{2}\PYG{p}{:,} \PYG{l+m+mi}{1}\PYG{p}{:}\PYG{o}{\PYGZhy{}}\PYG{l+m+mi}{1}\PYG{p}{]} \PYG{o}{+}
                             \PYG{n}{u1}\PYG{p}{[}\PYG{l+m+mi}{1}\PYG{p}{:}\PYG{o}{\PYGZhy{}}\PYG{l+m+mi}{1}\PYG{p}{,} \PYG{p}{:}\PYG{o}{\PYGZhy{}}\PYG{l+m+mi}{2}\PYG{p}{]} \PYG{o}{+}
                             \PYG{n}{u1}\PYG{p}{[}\PYG{l+m+mi}{1}\PYG{p}{:}\PYG{o}{\PYGZhy{}}\PYG{l+m+mi}{1}\PYG{p}{,} \PYG{l+m+mi}{2}\PYG{p}{:])}
    \PYG{n}{u1}\PYG{p}{,} \PYG{n}{u2} \PYG{o}{=} \PYG{n}{u2}\PYG{p}{,} \PYG{n}{u1}
\end{Verbatim}
\caption{Example of a 2D Laplace solver using CharmStencil}
\label{code:jacobi}
\end{figure}

\subsection{Frontend}

\subsubsection{Execution graph}

The CharmStencil frontend captures array operations as a parameterized Abstract Syntax Tree (AST). An AST is generated to represent the right-hand side of any expression that assigns to an array slice, and each AST stores the input and output arrays involved in that computation. The Python program is then represented as a Directed Acyclic Graph (DAG), where each node is an AST and edges represent true, anti, or output dependencies between the arrays of connected nodes.

For better frontend performance, dependencies are tracked at the array level rather than the index level. While this approach may add an unnecessary edge between two independent nodes, it guarantees that no required dependency is missed.

\begin{figure}
    \centering
    \begin{subfigure}[b]{.58\linewidth}
        \centering
        \includegraphics[width=\linewidth]{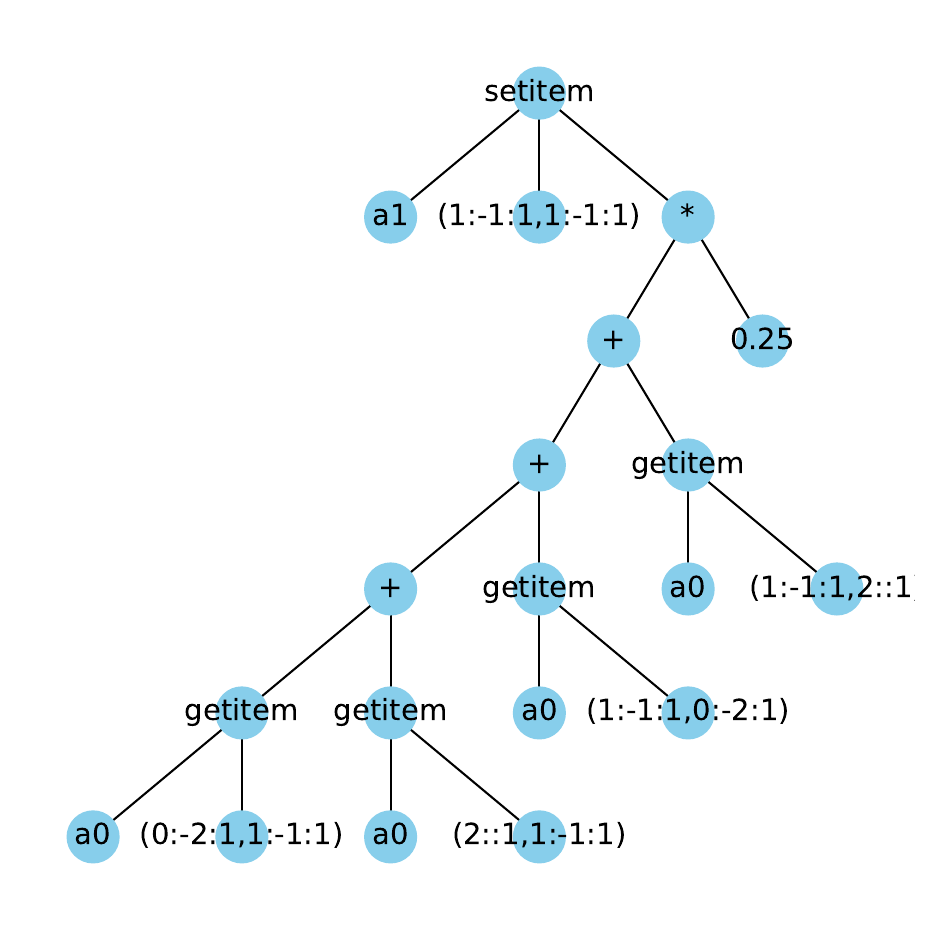}
        \caption{AST generated for the main loop expression (knl4 in the DAG)}
        \label{fig:jacobi:ast}
    \end{subfigure}    
    \begin{subfigure}[b]{.4\linewidth}
        \centering
        \includegraphics[width=\linewidth]{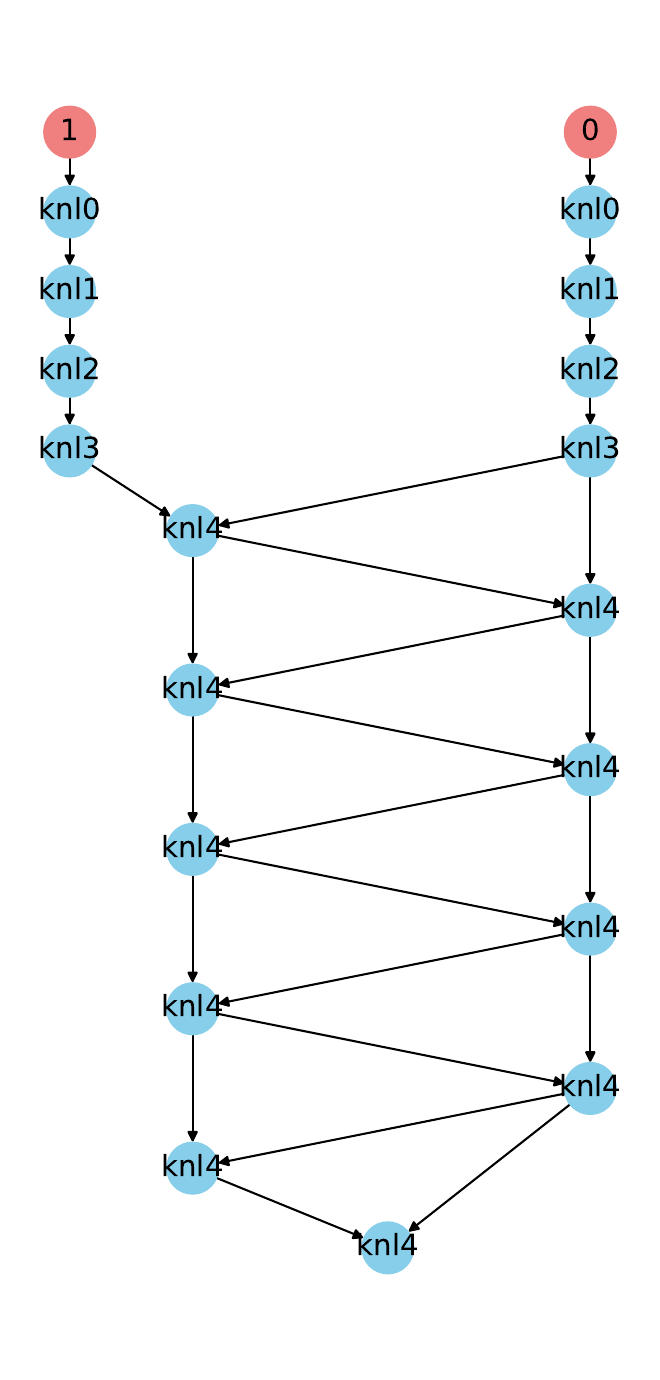}
        \caption{DAG generated for the example program~\ref{code:jacobi}}
        \label{fig:jacobi:dag}
    \end{subfigure}
    \caption{AST and DAG examples for 2D Laplace solver Example~\ref{code:jacobi}}
    \label{fig:jacobi}
\end{figure}

For example, Figure~\ref{fig:jacobi} shows the AST for the update expression inside the loop and the DAG for the entire program in Example 2. The red nodes in the DAG are the creation nodes for the arrays \texttt{u1} and \texttt{u2}. Because the AST is parameterized, all ten nodes representing the Jacobi relaxation iterations refer to the same AST, even when the arrays are swapped.

The DAG and its corresponding ASTs are sent to the backend when a \texttt{sync} function is called on the frontend, or when the DAG depth exceeds a user-configurable maximum.

\subsubsection{Node fusion}

In numerical methods, it is common for each grid point to have several fields that are updated at each timestep. In our abstraction, these fields are expressed as separate arrays of the same shape. Because we represent each assignment as a separate AST, cases where multiple fields are updated in a single iteration will result in multiple DAG nodes. These distinct ASTs are then compiled into separate CUDA kernels on the backend, which prevents potential data reuse and incurs additional kernel launch overheads.

To address this issue, we perform a node fusion pass over the DAG before sending it to the backend. We use two criteria to determine if nodes can be fused.

\paragraph{Output shape}
Fusing nodes with different output shapes would require the generated kernel to have complicated logic to map each thread to the correct output index for each expression. Moreover, if the sizes of the output arrays differ, it could create a load imbalance within the generated kernel. In stencil codes, the main computation loop typically iterates over the same set of grid points for all fields. Therefore, we restrict our node fusion criterion to nodes that have the same output shape.

\paragraph{Dependencies}
Two nodes can only be fused if they do not have any read-after-write or write-after-read dependencies between them. Since we express these dependencies as edges in our DAG, this criterion means that two nodes can be fused only if there is no path between them. Finding all pairs of nodes in a graph without a path is a computationally expensive problem~\cite{trans-closure}. However, in typical stencil codes, operations on different fields within an iteration are written consecutively. Thus, instead of checking all pairs of nodes, we use a heuristic that only attempts to fuse consecutive nodes in program order.

\subsection{Backend}

\subsubsection{Code generation}

The backend performs an analysis pass on each AST it receives from the frontend. During this pass, it tracks the memory access pattern of each argument to calculate the maximum offset between any input and output index. This information is stored as kernel metadata and is later used to determine the required ghost size for arrays. The backend also tracks data reuse, marking input arguments for use of shared memory if this optimization is enabled. For each kernel, the array arguments that are written to are tracked to assist with optimizing data movement, as discussed later in this section.

Following the analysis, the backend generates a CUDA kernel for each AST. The current code generator implementation assumes that each thread calculates at most one element of an output array slice. Consequently, the launch parameters are configured so that the total number of threads equals the size of the largest output slice. Each process compiles the generated CUDA kernel to the CUDA PTX IR separately, after which the kernel is dynamically loaded. This per-process compilation strategy was chosen to support systems without a shared filesystem, such as a cluster of AWS EC2 instances.

\subsubsection{Data layout}

On the backend, array data is tiled and distributed among chares. The total number of chares is determined by the number of PEs the backend is launched with and an over-decomposition factor set by the user at the frontend. All arrays are partitioned equally among all chares.

\begin{figure}
    \centering
    \includegraphics[width=.75\linewidth]{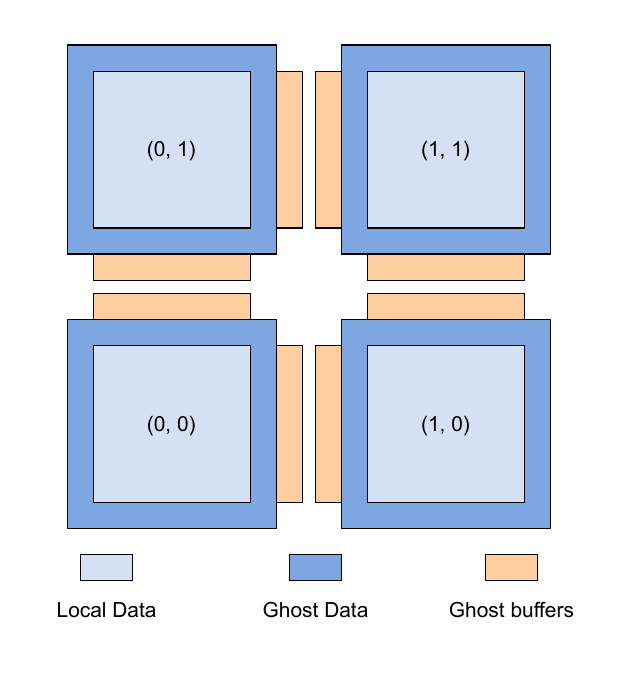}
    \caption{Data layout for $2 \times 2$ chare decomposition for an array with ghost depth 1}
    \label{fig:layout}
\end{figure}

Figure~\ref{fig:layout} shows the data layout for a $2 \times 2$ chare decomposition. Each chare's data structure has three components: the local data it owns, the ghost data from neighboring chares, and the send and receive buffers used for ghost exchanges. The size of the ghost region is inferred by taking the maximum ghost depth required by any kernel to which the array is passed as an argument.

\subsubsection{DAG execution}

Each chare creates a compute stream and a communication stream. All kernel launches are enqueued in the compute stream, while all memory copies and packing/unpacking kernels for ghost data are enqueued in the communication stream. Each chare also has two CUDA events: a compute event and a communication event. The compute event is recorded in the compute stream after a kernel launch, and the communication event is recorded in the communication stream after the latest memory copy or packing/unpacking kernel. These events are used to synchronize the compute and communication streams on each chare.

After the code for the ASTs is generated, the computation DAG is broadcast to all chares. Each chare executes the nodes in the DAG according to its specified dependencies. The execution of a DAG node occurs in two steps.

First, the ghost data for the input arrays is exchanged asynchronously. Before the packing kernels are called, a wait for the compute event is enqueued in the communication stream. This ensures that previous kernel executions are complete before the ghost data is copied into the send buffer. After the packing and unpacking kernels are called, the communication event is recorded.

Second, when the ghost data for all input arrays is available, a wait for the communication event is enqueued in the compute stream. This ensures that all ghost data has been copied into the send buffer and out of the receive buffer before the main kernel executes. The CUDA kernel is then enqueued in the compute stream. This asynchronous data exchange allows for the overlap of computation and communication between independent nodes in the DAG.

To determine which ghost exchanges are necessary for the execution of a DAG node, the array data structure maintains a generation epoch for both local and ghost data. When a kernel executes, the local generation epoch for the arrays that are written to is incremented. When the ghost data for an array is exchanged, its ghost generation epoch is updated to match the local generation epoch. Before a DAG node is executed, the kernel metadata is checked to identify arrays with a non-zero ghost depth. A ghost exchange is performed only for these arrays and only if their local and ghost generation epochs do not match.
%\subsection{Execution graph}

\subsection{Limitations \& Future Work}

While NumPy's slicing notation is a powerful way to express computations on structured grids, it restricts applications to those where computations can be expressed as operations on array subviews. Consequently, irregular algorithms—such as particle methods and unstructured-grid solvers—cannot be expressed using this abstraction.

Our ghost-region-based remote data transfer scheme places an additional restriction on the application's dependence relation: a grid point $i$ must not depend on a grid point $i \pm k$ where $k$ is greater than the width of a chare, as remote data is only transferred between neighboring chares. Furthermore, our current implementation does not optimize ghost data transfers involving strided views, such as those found in geometric multigrid methods. We plan to address these limitations in future work by implementing a more general remote data transfer mechanism.

\section{Resource Elasticity}

Resource elasticity requires support from both the job scheduler and the parallel programming model. The job scheduler must be able to dynamically change a job's allocation, while the programming model must be able to adapt to this new allocation. The specifics of job scheduler support are outside the scope of this work. In this paper, we focus on providing elasticity within the programming model.

In Charm++, resource elasticity is supported using its shrink/expand functionality. Previously, this support was limited to CPUs and the \texttt{netlrts} communication layer. To enable resource elasticity in our abstraction, we extended the shrink/expand functionality in Charm++ to support GPUs and generalized the implementation to work with any communication layer. Our abstraction uses the UCX layer, which supports GPUDirect communication across devices. Extending shrink/expand support to GPUs involved two major contributions.

\subsubsection{Migration of chares}

Chare migration is an essential part of the shrink/expand process. During a shrink operation, chares on PEs designated for removal are migrated to other PEs. During an expand operation, chares are migrated to new PEs to rebalance the load. The migratability of chares is enabled by the PUP framework in Charm++, which we extended to support the migration of GPU-resident data using GPUDirect RDMA.

\subsubsection{Checkpoint and restart}

During a shrink/expand operation, local chare data is checkpointed to Linux shared memory, the application is restarted with the new number of processes, and the checkpoint data is restored. With GPU-resident data, this approach requires that data persists on the GPU through an application restart. However, the lifetime of data on a GPU is tied to the CUDA context that owns it, and a CUDA context cannot persist across a process restart. One solution is to copy all data from the device to the host and checkpoint it to Linux shared memory. Then, upon restart, the checkpoint data is restored and copied back to the GPU. However, host-to-device transfers are typically expensive due to the low bandwidth of the PCIe bus.

\begin{figure}
    \centering
    \includegraphics[width=\linewidth]{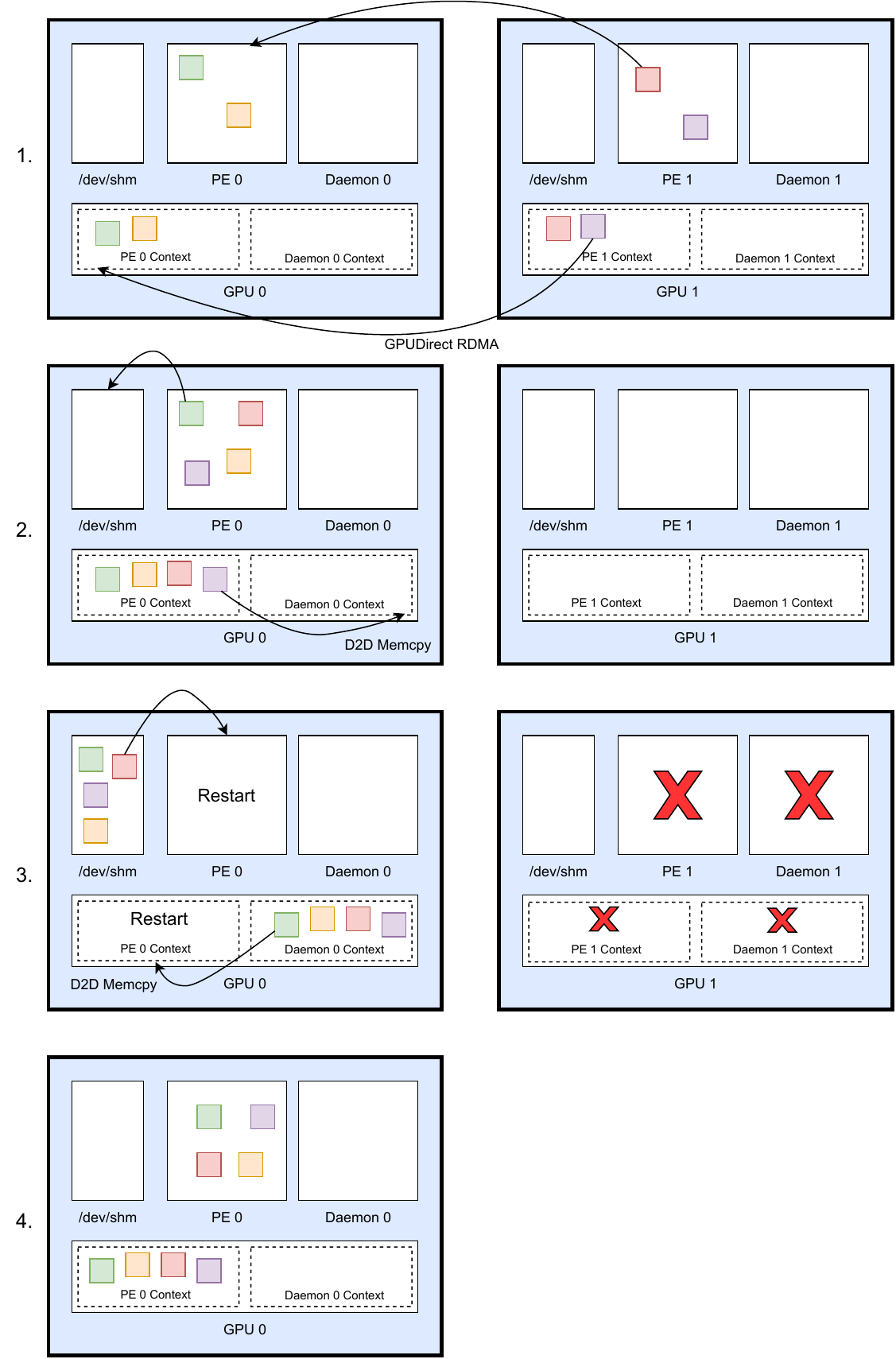}
    \caption{Shrinking from 2 PEs on 2 nodes with 1 GPU each, to 1 PE}
    \label{fig:shrinkexpand-gpu}
\end{figure}

To avoid this costly data movement between the host and device, we launch a separate memory daemon process for every GPU at application startup. When the application is signaled to rescale, the GPU-resident data pointers on all chares are sent to the corresponding memory daemons using CUDA IPC and a named pipe. The memory daemons copy the data into a buffer they own and return an allocation ID to the Charm++ runtime. This allocation ID is written into the shared memory checkpoint, and the application is restarted while the memory daemons remain active. During the application's restore stage, the Charm++ runtime retrieves the device pointer corresponding to the allocation ID and copies the data back into a device buffer allocated by the Charm++ runtime. The corresponding data on the memory daemon is then freed. This scheme replaces expensive device-to-host data movement with more efficient device-to-device copies. Figure~\ref{fig:shrinkexpand-gpu} shows an end-to-end illustration of shrinking from two PEs on two nodes to one PE. The rescaling functionality is exposed to the user in CharmStencil through a \texttt{rescale} call.

\section{Results}

For our experiments we used the TACC Vista supercomputer. Each node on Vista has a single NVIDIA GH200 GPU with an Infiniband interconnect. We used a \texttt{ucx-linux-arm8-cuda} non-SMP Charm++ build.

\subsection{Pipelined execution performance}

CharmStencil executes the frontend and backend in a pipelined fashion to hide the overhead of the Python frontend. To study how effectively this model hides the overhead, we evaluated the performance of the Laplace solver from Figure~\ref{code:jacobi} with a varying DAG depth. The experiment used a grid size of $16k \times 16k$ per GPU across 4 GPUs on 4 nodes and ran for 10,000 iterations.

\begin{figure}
    \centering
    \includegraphics[width=.75\linewidth]{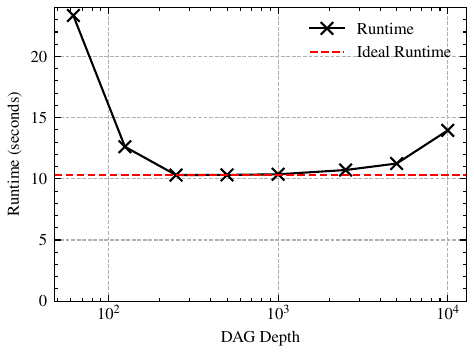}
    \caption{Performance of a Laplace solver with varying DAG depth}
    \label{fig:pipeline-perf}
\end{figure}

Figure~\ref{fig:pipeline-perf} shows the runtime with varying DAG depths compared to the ideal runtime (i.e., without any Python overhead). For very small DAG depths, the runtime is high due to frequent communication between the frontend and the backend. Conversely, a large DAG depth reduces this communication overhead but also reduces the overlap between frontend and backend execution. In Figure~\ref{fig:pipeline-perf}, a DAG depth of 10,000 results in the frontend executing the entire program before any operations are sent to the backend. Consequently, the total runtime includes the full Python overhead from the frontend, which was 3.7s in this example.
Our results show that for a large range of DAG depths between these two extremes, our execution model effectively hides the Python overhead.

\subsection{Scaling performance}

We measured weak scaling performance on two applications - a Laplace equation solver from figure~\ref{code:jacobi}, and a more complex CFD benchmark for simulating a cavity flow using the Navier-Stokes equations~\cite{Barba2019}.
We compared our results with those of cuPyNumeric and Devito. For the Laplace equation solver, we also compared our results with a handwritten Charm++ and CUDA implementation.
cuPyNumeric is a drop-in NumPy replacement library based on the Legion runtime system~\cite{cupynumeric} that supports execution on multi-node, multi-GPU systems using UCX and GASNet. For our experiments, we used a version of cuPyNumeric built with UCX. Devito is a high-level, Python-based DSL for stencil computations~\cite{devito-compiler}. The open-source version of Devito supports GPU execution using OpenACC and can run on systems with multiple GPUs on a single node, but it does not support multi-node execution. We measured the runtime for Devito with the \texttt{advanced} optimization mode.

\begin{figure*}[ht]
    \centering
    \begin{subfigure}[b]{.36\linewidth}
        \centering
        \includegraphics[width=\linewidth]{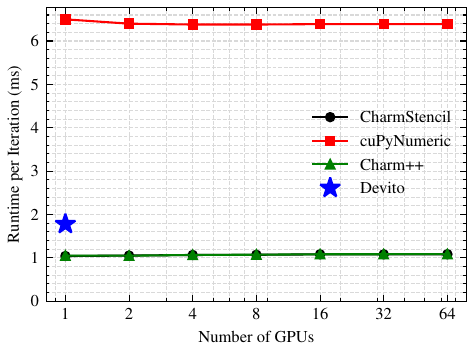}
        \caption{Laplace equation}
        \label{fig:scaling:laplace}
    \end{subfigure}
    \begin{subfigure}[b]{.36\linewidth}
        \centering
        \includegraphics[width=\linewidth]{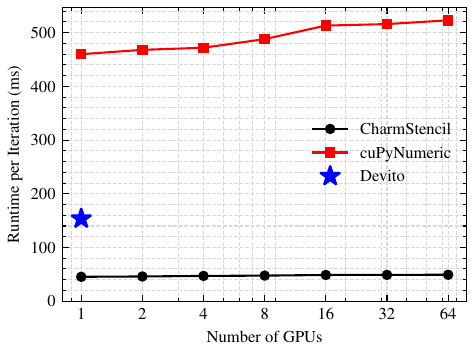}
        \caption{Cavity flow}
        \label{fig:scaling:cfd}
    \end{subfigure}
    \caption{Weak scaling results}
    \label{fig:scaling}
\end{figure*}

For the Laplace equation solver, we used a $16k \times 16k$ grid per GPU with 1000 iterations. For the cavity flow solver, we used a grid size of $16k \times 16k$ per GPU and 50 iterations. Figure~\ref{fig:scaling} shows the weak scaling results for both benchmarks. Since Devito only supports single-node execution and the Vista supercomputer has one GPU per node, Devito's results are limited to a single GPU.

The results show that CharmStencil is faster than cuPyNumeric by a factor of 6 for the Laplace solver and a factor of 10 for the cavity flow solver, exhibiting near-perfect weak scaling. This speedup is likely because of the lack of kernel fusion in the open-source version of cuPyNumeric. On a single GPU, CharmStencil performs 1.7x better than Devito for the Laplace solver and 3.2x better for the cavity flow benchmark.
We also see that CharmStencil performs on par with a handwritten Charm++ and CUDA implementation of the Laplace equation solver while reducing the code size from 830 to just 30 LoC.

\subsection{Rescaling overhead}

Rescaling in Charm++ has four sources of overhead:

\begin{enumerate}
\item Load balancing: When shrinking, chares and their corresponding GPU-resident data are moved away from nodes that will be removed. When expanding, a load balancing step migrates chares to newly added nodes to rebalance the workload.
\item Checkpoint: The chare's CPU data is checkpointed to Linux shared memory, while the GPU-resident data is copied to the corresponding daemon process before the application restarts.
\item Restart: The application restart itself incurs a startup overhead.
\item Restore: The chare's CPU data is read from Linux shared memory, and the GPU-resident data is retrieved and copied from the daemon process back to the Charm++ PE after the restart.
\end{enumerate}

\begin{figure*}
    \centering
    \begin{subfigure}[b]{.32\linewidth}
        \centering
        \includegraphics[width=\linewidth]{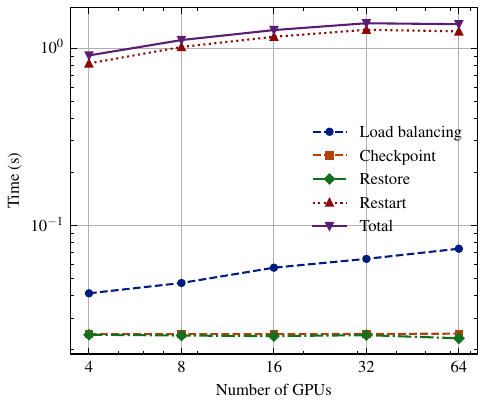}
        \caption{Shrinking to half the number of nodes. The x-axis shows the number of nodes before shrinking}
        \label{fig:overhead:shrink}
    \end{subfigure}
    \hfill
    \begin{subfigure}[b]{.32\linewidth}
        \centering
        \includegraphics[width=\linewidth]{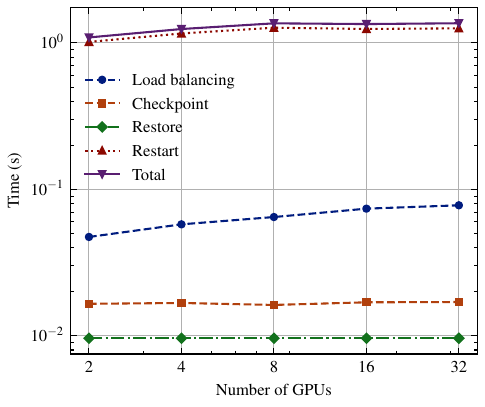}
        \caption{Expanding to double the number of nodes. The x-axis shows the number of nodes before expanding}
        \label{fig:overhead:expand}
    \end{subfigure}
    \hfill
    \begin{subfigure}[b]{.32\linewidth}
        \centering
        \includegraphics[width=\linewidth]{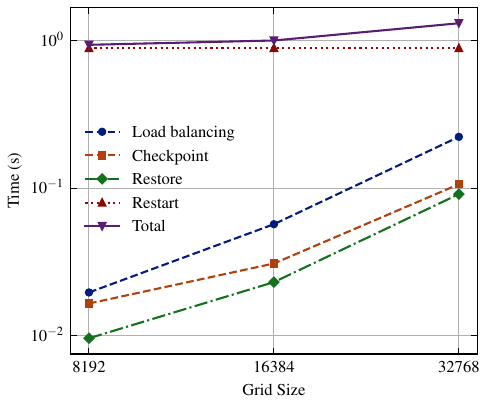}
        \caption{Shrinking from 16 to 8 nodes for different problem sizes. Grid size is the size of one dimension of the 2D grid per GPU}
        \label{fig:overhead:size}
    \end{subfigure}
    \caption{Contribution of different stages of rescaling to the total rescaling overhead}
    \label{fig:overhead}
\end{figure*}

To study the contribution of each source to the total overhead, we measured each stage of rescaling by shrinking and expanding a CharmStencil application within a SLURM allocation. We first studied the overhead on a varying number of nodes with a constant problem size per GPU. Figures~\ref{fig:overhead:shrink} and~\ref{fig:overhead:expand} show the overheads from shrinking to half the number of nodes and expanding to double the number of nodes, respectively. For the shrink tests, we used a data size of 1GB per GPU; for the expand tests, we used 2GB per GPU. In both cases, the results show that the overhead was dominated by the application restart time, with the total overhead on the order of one second.

To study the effect of data size on the rescaling overhead, we measured the cost of shrinking from 16 to 8 nodes with data sizes varying from 250MB to 4GB per GPU. We observed that the checkpoint, restore, and load balancing overheads scale linearly with the data size; however, the total overhead remains on the order of one second.

\subsection{Performance with rescaling}

\begin{figure}
    \centering
    \includegraphics[width=.75\linewidth]{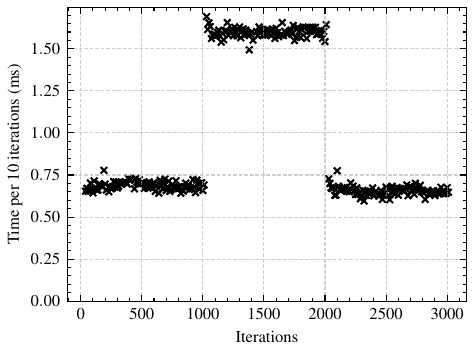}
    \caption{Time per 10 iterations of a Laplace solver with 2 rescaling calls, from 8 to 4 nodes, and back from 4 to 8 nodes}
    \label{fig:timeline}
\end{figure}

To demonstrate the resource elasticity of CharmStencil, we used the Laplace solver with a $32k \times 32k$ grid distributed across 8 nodes. The application ran for 1000 iterations, after which we called \texttt{rescale} to scale the job down to 4 nodes. After another 1000 iterations, a second \texttt{rescale} call expanded the job back to 8 nodes. The runtime was measured every 10 iterations on the backend, as shown in Figure~\ref{fig:timeline}. As expected, the time per iteration doubles after the first rescale event because the application is running on half the number of GPUs. Following the second \texttt{rescale} call, the time per iteration returns to its original value as the application expands back to its initial 8-GPU allocation.

\section{Related Work}

Several DSLs have been developed for stencil computations, broadly classified into two categories: high-level stencil abstractions and general NumPy drop-in replacements.

\subsection{NumPy drop-in replacement DSLs}

CuPy is a direct NumPy replacement for single GPUs, leveraging NVIDIA's highly optimized libraries (cuBLAS, cuDNN, etc.) for array operations~\cite{cupy}. cuPyNumeric is a distributed replacement that internally uses CuPy for array operations and the Legion runtime for managing distributed execution~\cite{cupynumeric}. cuPyNumeric shows excellent weak scaling performance on various benchmarks, including CFD applications, using optimizations like control replication to minimize scheduling overhead and kernel fusion to improve data reuse.

\subsection{Specialized stencil DSLs}

Devito is a high-level, Python-based symbolic DSL for stencil computations~\cite{devito-compiler}. Rather than the NumPy API, Devito uses a higher-level mathematical representation of equations, which it internally discretizes and JIT-compiles. Devito targets shared-memory (OpenMP), distributed-memory (MPI), and GPU (OpenACC) systems.

OPS is a C/C++ and Fortran-based DSL for computations on structured meshes~\cite{OPS}. Users write computations using data blocks and parallel loop constructs; OPS then uses a source-to-source translator to generate high-performance parallel code for shared-memory CPUs, distributed-memory CPUs, and GPUs.

\section{Conclusion}

In this paper, we present CharmStencil, a distributed, adaptive GPU stencil abstraction for Charm++. Existing stencil DSLs often force a trade-off between NumPy-like syntax—requiring significant porting effort—and performance. CharmStencil addresses this gap by providing a familiar NumPy-like frontend with a highly optimized backend.

A key challenge in Python HPC is interpreter overhead. We show that CharmStencil effectively hides this overhead using a pipelined execution model. Furthermore, our performance evaluations demonstrate that CharmStencil significantly outperforms both a generalized NumPy replacement and a specialized stencil abstraction, and performs on par with a handwritten Charm++ CUDA implementation.

Resource elasticity is increasingly becoming a necessity on modern supercomputing systems and cloud platforms. We extended the rescaling support in Charm++ to GPUs, incorporating elasticity into the CharmTyles framework. CharmStencil, built on CharmTyles, efficiently supports this feature. We demonstrated this by dynamically scaling an application, measuring the overhead of rescaling at approximately one second on a SLURM-based supercomputer.

\section{Acknowledgments}

This work was funded by the IBM-Illinois Discovery Accelerator Institute (IIDAI).

\bibliographystyle{plain} % We choose the "plain" reference style
\bibliography{citations, group} % Entries are in the refs.bib file

\end{document}